\documentclass[superscriptaddress,aps,pra,twocolumn,showpacs,preprintnumbers,amsmath,amssymb,floatfix,groupedaddress]{revtex4-1}
\usepackage{graphicx}
\usepackage{dcolumn}
\usepackage{bm}
\usepackage{color}
\usepackage{psfrag}
\begin{document}
%



\title{Controllable steep dispersion with gain in a four-level \emph{N}-scheme with four-wave mixing}

\author{Nathaniel B. Phillips$^{a}$, Irina Novikova$^{a}$, Eugeniy E. Mikhailov$^{a}$, Dmitry Budker$^{b,c}$, Simon Rochester$^{b}$\\
\vspace{6pt}
$^{a}${\em{Department of Physics, The College of William and Mary, Williamsburg, VA 23185, USA}}; $^{b}${\em{Rochester Scientific, LLC, El Cerrito, CA, 94530, USA}};
$^{c}${\em{Department of Physics, University of California, Berkeley, CA 94720, USA}}\\\vspace{6pt}
\date{\today}}

\begin{abstract}
We present a theoretical analysis of the propagation of light
pulses through a medium of four-level atoms,  with two strong
pump fields and a weak signal field in an $N$-scheme
arrangement. We show that the generation of four-wave mixing
has a profound effect on the signal field group velocity and
absorption, allowing the signal field propagation to be tuned
from superluminal to slow light regimes with amplification.

\end{abstract}
\maketitle

\section{Introduction}

\begin{figure}[htp]
\centering
\includegraphics[width=0.8\columnwidth]{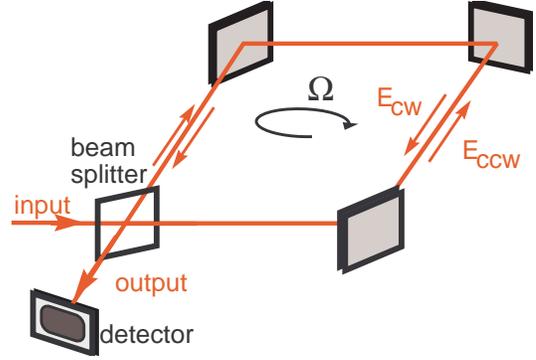}%
\caption{Generic schematic of a generic optical gyroscope based on
the Sagnac effect.}
\label{optgyro.fig}%
\end{figure}

Precise rotation sensors are critical components for
stabilization, navigation, and targeting applications. At the
moment, the most sensitive commercial devices are optical
gyroscopes based on the Sagnac effect~\cite{sagnac}. Such a
device consists of a ring interferometer with two
counter-propagating light waves, as shown in
Fig.~\ref{optgyro.fig}. The rotation of such an interferometer
results in a phase difference between the two optical fields
proportional to the magnitude of the rotational angular
velocity $\vec{\Omega}$:
\begin{equation}\label{sagnac}
  \Delta\phi=\frac{4\pi\omega}{c^2}\vec{A}\cdot\vec{\Omega},
\end{equation}
where $\omega$ is the light angular frequency, $c$ is the speed
of light, and $\vec{A}$ is the area of the optical loop. Most
successful realizations to date are fiber-optics gyroscopes, in
which the interferometer ring is formed by a loop of an optical
fiber. The sensitivity of such an interferometer is usually
boosted by using a large number $N$ of loops that increase the
effective area in Eq.~\eqref{sagnac} by a factor of $N$. The Sagnac
phase shift can then be measured directly from the interference
of the two counter-propagating waves at the output, or by
monitoring the resulting frequency difference between
corresponding counter-propagating modes of the interferometer
cavity. In either case, the reciprocity of light propagation
dramatically reduces effects of environmental factors
(temperature, vibrations, etc.), and ensures high reliability.
As a result, the sensitivity of state-of-the-art compact
fiber-optic gyroscopes has reached the shot-noise-limited value
of
$10^{-7}$--$10^{-8}~\mathrm{rad/s/\sqrt{Hz}}$~\cite{optgyrosens},
while large-area laser gyroscopes have achieved even greater
sensitivities, on the order of
$10^{-10}~\mathrm{rad/s/\sqrt{Hz}}$~\cite{lasergyro}.

Similar sensitivity has been also achieved with matter-based
Sagnac interferometers. In this case, the rotation-induced phase
equation may be written as
\begin{equation}\label{sagnac_matter}
    \Delta\phi= \frac{4\pi}{\lambda_{\rm dB}{v}}\vec{A}\cdot\vec{\Omega},
\end{equation}
where ${v}$ and $\lambda_{\rm dB}=2\pi\hbar/(mv)$ represent the
average velocity and the de Broglie wavelength of the massive
particles, respectively. Here, the advantage gained by the use
of massive particles ($mc^2\gg\hbar\omega$) is offset by the
much smaller effective area compared to fiber-optics devices,
resulting in similar performance~\cite{kasevich00}.

Recent demonstrations of slow light pulse propagation in
coherent optical media stirred active debate on the possibility
of using slow-light pulses to enhance the Sagnac effect. It
was quickly established that neither large positive (``slow
light'') nor negative (``fast light'') dispersion has a
\emph{direct} influence on the magnitude of the Sagnac phase
shift in Eq.~(\ref{sagnac})~\cite{malykin00}.

Nevertheless, it still seems to be possible to take advantage
of a large group index to enhance gyroscopic performance. For
example, the output signal of a rotating interferometer with a
highly dispersive slow-light medium can be enhanced by its
differential response to opposite Sagnac phase shifts of two
counter-propagating light waves~\cite{PengLiXu07}. A modest
factor-of-$2.5$ enhancement of the observed phase difference has been
recently demonstrated in a slow light fiber ring~\cite{Yuan10},
and a more significant enhancement (up to a factor of 200) is
predicted in certain coupled resonator
structures~\cite{PengLiXu07}.

Even more dramatic improvements are predicted for the measurement
of the Sagnac-effect-induced mode splitting in an active ring
cavity with strong negative dispersion~\cite{shahriarPRA07}.
Calculations have shown that the resulting frequency difference
between two counter-propagating modes is inversely proportional
to the group index, and thus nominally diverges for $n_g=0$
(i.e., for $n\simeq -\omega \frac{\partial
n}{\partial\omega}$)~\cite{shahriarPRA07,shahriarOC08}. While
this divergence disappears after correcting for higher-order
nonlinear effects, a $10^6$ improvement in gyroscope
sensitivity should still be possible.

The current status of these debates shows that while strong
positive or negative optical dispersion may indeed be capable
of dramatic improvements in optical gyroscope
performance, there is no clear winning approach. Thus, an
atomic system that can be easily reconfigured to exhibit either
strong positive or strong negative dispersion is an ideal
candidate for the development of such a new generation of
advanced optical gyroscopes. In the last decade, controllable
manipulations of the group velocity of light have been
demonstrated in a wide range of
systems~\cite{Boyd2002,Boyd2009Controlling}. Nonetheless, atomic
systems with long-lived spin coherences still provide the
highest values of group index for both slow and fast light
regimes~\cite{Akulshin2010}. In such atomic systems, the group
velocity for a probe optical field can be widely tuned by
adjusting parameters of a strong control field that provides
strong coupling of the probe optical field to a collective
atomic spin state~\cite{lukin03rmp}.

\begin{figure}[htp]
\includegraphics[width=0.6\columnwidth]{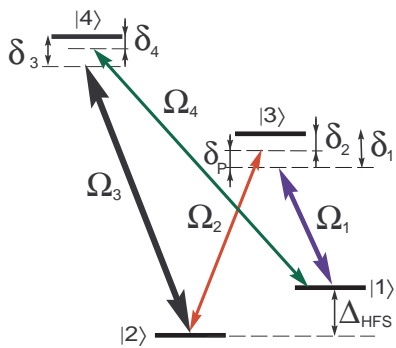}
\caption{Schematic for four optical fields interacting with
four-level atoms in an \emph{N}-configuration.}
\label{Nlevels.fig}%
\end{figure}

An ideal test system for the development of a new type of
optical gyroscope with improved rotational sensitivity should
have a dispersion that can be continuously controlled in the
widest possible range---from the highest positive group index to
the highest negative group index---with minimal changes in the
experimental arrangement. While several interaction schemes are
capable of such wide tunability~\cite{akulshin03,mikhailov04},
a so-called \emph{N}-scheme has recently emerged as a promising
candidate~\cite{harris98,YZhu04,Ham09,narducciJMO11,Narducci10}. A
possible realization of an \emph{N}-scheme is formed by three
optical fields interacting with four-level atoms in the
arrangement shown in Fig.~\ref{Nlevels.fig}. In the absence of
the control field, the two resonant optical fields
$\Omega_{1,2}$ form a regular $\Lambda$ system exhibiting EIT
and slow light~\cite{lukin03rmp}. The interaction of the atoms with the second
strong control field $\Omega_3$ splits this single EIT peak
into two, separated by a narrow enhanced-absorption peak. This
spectral region exhibits a fast-light effect, desired for
gyroscope performance enhancement. However, this fast-light
regime cannot be directly utilized in the proposed active
enhanced-sensitivity optical gyroscope due to its unavoidable
high optical losses.

In this manuscript, we provide an extended treatment of the
four-level \emph{N}-scheme that includes the possibility of
four-wave mixing (FWM) by allowing optical transitions (and
spontaneous decay) between states $|4\rangle$ and $|1\rangle$.
The associated FWM gain modifies the transmission of the probe
field~\cite{FleischhakerPRA08,LettFWM12}, and provides a smooth
switch between slow- and fast-light regimes by varying the
strength of one of the pump fields ($\Omega_3$).

\section{Slow and Fast light in a four-level \emph{N}-scheme}

The evolution of a four-level \emph{N}-system, shown in
Fig.~\ref{Nlevels.fig}, can be described under the
rotating-wave approximation by the following Hamiltonian:
\begin{equation}
\frac{\hat{H}}{ i \hbar} = \left(
\begin{array}{cccc}
 0 & 0 & - \frac{1}{2} e^{-i \phi_1}\Omega_1 & -\frac{1}{2} e^{-i \phi_4}\Omega_4 \\
 0 & -\delta_1+\delta_2 & -\frac{1}{2} e^{-i \phi_2}\Omega_2 & -\frac{1}{2} e^{-i \phi_3}\Omega_3 \\
 -\frac{1}{2} e^{i \phi_1}\Omega_1 & -\frac{1}{2} e^{i \phi_2}\Omega_2 & -\delta_1 & 0 \\
 -\frac{1}{2} e^{i \phi_4}\Omega_4 & -\frac{1}{2} e^{i \phi_3}\Omega_3 & 0 & -\delta_1+\delta_2-\delta_3
\end{array}
\right), \label{hamiltonian}
\end{equation}
where $\Omega_i$ and $\phi_i$ are the Rabi frequencies and
phases of the corresponding optical fields, respectively, and
$\delta_i$ are their detunings from the corresponding optical
transitions, as shown in Fig.~\ref{Nlevels.fig}. Here we have assumed
the four-photon resonance condition $-\delta_1+\delta_2
-\delta_3+\delta_4=0$, as well as the phase-matching condition
on the optical wavenumbers $k_i$, $-k_1+k_2-k_3+k_4=0$, which
results in the elimination of the explicit time and space
dependence from the Hamiltonian~\cite{mahmoudiPRA06}. The
four-photon resonance condition is automatically satisfied in
the situation that we will primarily consider, in which the
Stokes field $\Omega_4$ is spontaneously generated.

\begin{figure}[htp]
\includegraphics[width=1.0\columnwidth]{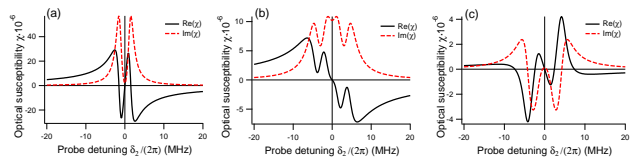}%
\caption{Real and imaginary parts of the probe-field susceptibility for various interaction
configurations: \emph{(a)} only one control field $\Omega_1= (2
\pi)\, 3$~MHz is on (standard EIT regime); \emph{(b)} both
control fields $\Omega_1= (2 \pi)\, 3$~MHz and $\Omega_3= (2
\pi)\, 6$~MHz are present, but no radiative transition between
states $|4\rangle$ and $|1\rangle$ is allowed (standard
\emph{N}-scheme); \emph{(c)} both control fields $\Omega_1= (2
\pi)\, 3$~MHz and $\Omega_3= (2 \pi)\, 6$~MHz are present, and
both excited states have equal decay rates into each of the ground
states. For all graphs the excited state decay rates are
$\gamma_{3} = \gamma_{4} = (2 \pi)\, 3$~MHz, the ground-state relaxation rate is $(2 \pi)\, 0.01$~MHz, we assume equal branching ratios for all optical
transitions, and both pump fields are resonant with corresponding
optical transitions. }
\label{Nscheme_comp.fig}%
\end{figure}

The ability to control the dispersion of the probe field
$\Omega_2$ by adjusting the intensities of two strong control
fields $\Omega_1$ and $\Omega_3$ is illustrated in
Fig.~\ref{Nscheme_comp.fig}, obtained by numerically solving
the evolution equations obtained from the above Hamiltonian for
the steady-state condition. Figure~\ref{Nscheme_comp.fig}(a) shows a traditional EIT
regime, with a moderately strong first control field $\Omega_1
= (2 \pi)\, 3$~MHz  and the second control field $\Omega_3$
turned off. As expected, we observe a dip in the absorption
spectrum (dashed line) and steep, positive, linear dispersion
of the refractive index (solid line) near zero two-photon
detuning $\delta_P=\delta_2-\delta_1=0$, between two absorption
peaks corresponding to the Autler-Townes splitting of the
excited state by the strong control field.
%
%
Figure~\ref{Nscheme_comp.fig}(b) depicts the situation in which
the atoms interact with both strong control fields $\Omega_1$
and $\Omega_3$ in a standard \emph{N}-configuration, in which
optical transition from state $|4\rangle$ to $|1\rangle$ is
not allowed by selection rules. In this case, the
spectrum consists of four partially-resolved absorption
resonances, which can be interpreted as unequal Autler-Townes
splittings of the states $|2\rangle$ and $|3\rangle$ by the
control fields of different intensities $\Omega_1 = (2 \pi)\,
3$~MHz and $\Omega_3 = (2 \pi)\, 6$~MHz. Even though there are
several spectral regions in which steep anomalous dispersion is
realized, all of them occur in conjunction with enhanced
absorption.

Finally, Figure~\ref{Nscheme_comp.fig}(c) shows that the
situation is quite different if optical transitions are allowed
from both excited states to each of the ground states.
In this case, the four-wave mixing process in a double-$\Lambda$ system is possible, and
it is enhanced through the long-lived spin coherence between states $|1\rangle$ and $|2\rangle$~\cite{lukin03rmp,phillipsJMO09,LettFWM12}. As a result, a new optical Stokes field $\Omega_4$ is efficiently generated, and the probe-field spectrum consists of two antisymmetric Raman resonances, with gain regions
at both positive and negative probe-field detunings. For properly chosen intensities of the two control fields, it is possible to adjust the frequency splitting and widths of these peaks to achieve a negatively-sloped refractive index for the probe field near the zero two-photon
detuning $\delta_P=0$, while the its gain drops to zero between
the two gain peaks. Thus, the probe field
experiences minimal absorption or gain for frequencies near the
two-photon resonance, which are the desired characteristics of
an atomic medium for gyroscope enhancement.

\begin{figure}[htp]
\includegraphics[width=1.0\columnwidth]{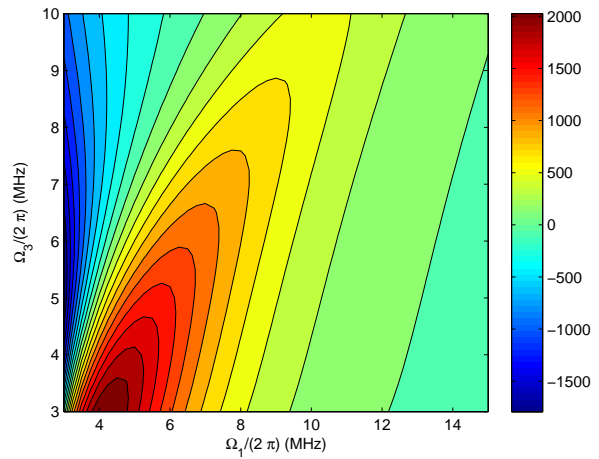}%
\caption{ Optimization of the probe-field group index $n_g$ on the two-photon resonance,  as
function of both control fields' strengths. For this calculation, we assumed $1$~cm-long atomic medium of $10^{9}$~cm Rb density; the rest of the experimental
parameters are the same as in Fig.~\ref{Nscheme_comp.fig}. Zero probe-field absorption is predicted
for the shown range of control fields' Rabi frequencies.}
\label{dispersionmap.fig}%
\end{figure}

From this picture, it is clear that optimization of the control field
intensities allows for smooth tuning of the probe field's dispersion from slow to fast light regimes by changing the frequency shift and shape of the Raman peaks. To find the optimal operational parameters
numerically, we compute the spectrum of the probe field
$\Omega_2$ for the range of the control fields' Rabi
frequencies and calculate dispersion at the zero two-photon
resonance. The results are shown in
Fig.~\ref{dispersionmap.fig}. One can see that depending on the
ratio between two control fields, the probe experiences either slow light (when the two gain peaks for positive and negative two-photon detuning are not resolved and form a single gain peak), or fast light (when the two peak are farther apart, forming a distinct dip between them). When both fields are very strong, the Raman resonances are shifted too far from the origin, leading to flat dispersion.
From this analysis we have identified
$\Omega_1 = (2 \pi)\, 3$~MHz and $\Omega_3 = (2 \pi)\, 6$~MHz
as suitable values for producing the desired lossless
fast-light behavior.

\begin{figure}
\includegraphics[width=1.0\columnwidth]{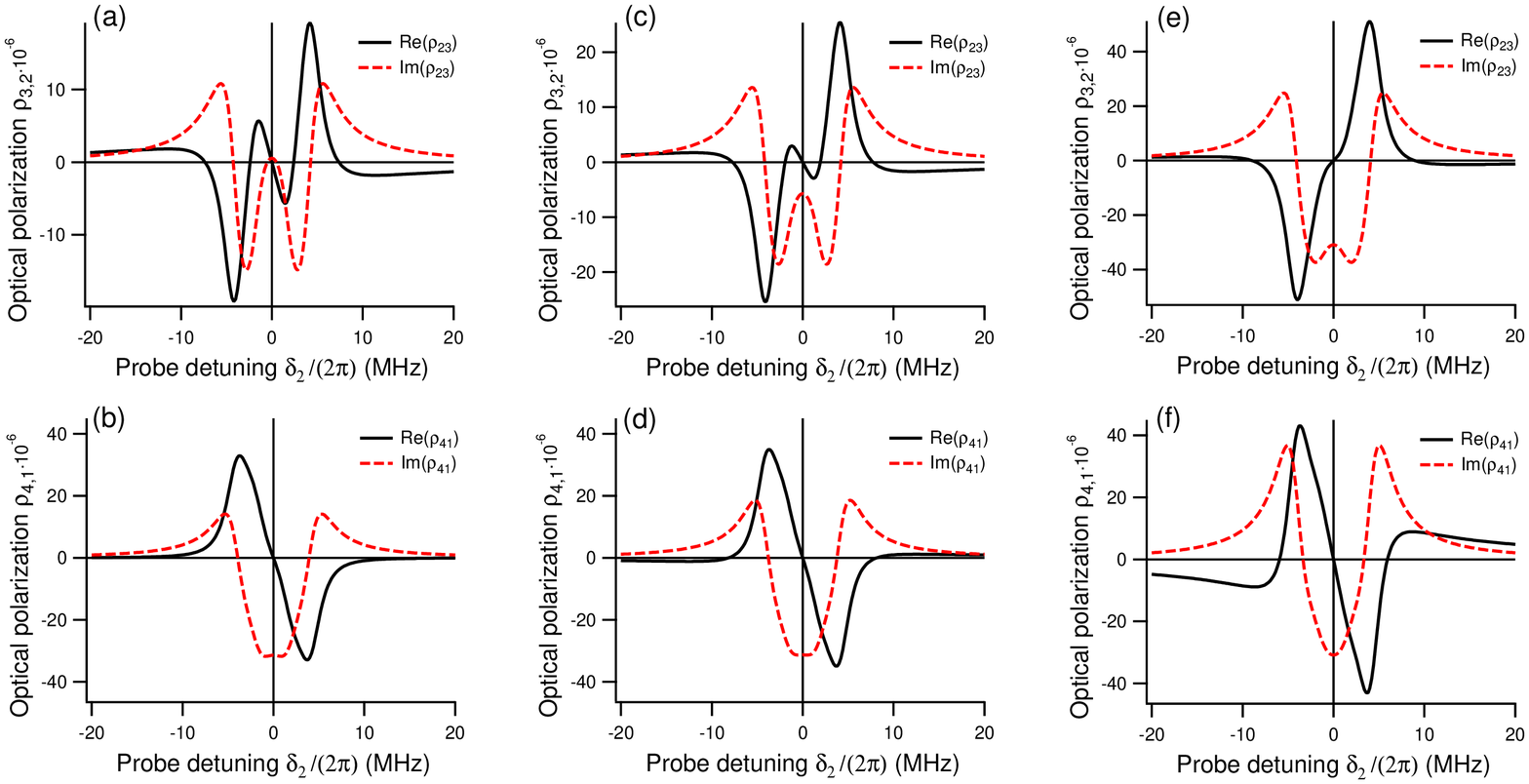}%
\caption{\emph{Top row}: Real and imaginary parts of atomic polarization $\rho_{3,2}$, proportional to the refractive index and absorption for the probe
field $\Omega_2$ for for various strengths of the Stokes field. \emph{Bottom row}: Same for real and imaginary parts of atomic polarization $\rho_{4,1}$ for the Stokes field  $\Omega_4$. Pump fields are
$\Omega_1 = (2 \pi)\, 3$~MHz and $\Omega_3 = (2 \pi)\, 6$~MHz. }
\label{FWManalysis.fig}%
\end{figure}

Under the four-wave mixing condition, the spontaneously
generated Stokes field $\Omega_4$ experiences strong gain, and
thus its intensity increases as it propagates through the medium.
Moreover, its presence has a strong effect on the probe field amplitude due to their mutual coupling through the atomic spin coherence, even though both probe and Stokes fields remain significantly weaker than either control field.
In Fig.~\ref{FWManalysis.fig}, we plot the real and imaginary
parts of optical polarizations for both the probe (top) and
Stokes (bottom) fields, under conditions corresponding to
different points along the optical path through the atomic
medium. The left column represents the entrance of the vapor
cell, where only the probe field is present, and $\Omega_4=0$
since it is not yet generated. Under these conditions,
$\Omega_4$ experiences strong gain, which leads to its spontaneous
generation. The Stokes field is generated at the frequency that
satisfies the four-photon resonance condition---any variation
in the probe two-photon detuning $\delta_P=\delta_2-\delta_1$
is matched by the corresponding change in the Stokes field
two-photon detuning $\delta_S=\delta_4-\delta_3 = - \delta_P$.

As the unattenuated probe light and generated Stokes field
propagate along the cell, the increasing strength of $\Omega_4$
starts affecting the propagation of the probe field through the
FWM coupling. In particular, the negatively-sloped refractive
index is somewhat flattened out, due to appearance of a small
amount of gain [Fig.~\ref{FWManalysis.fig}(c)]. Farther along
the cell, the probe field experiences stronger gain, but the
dispersion switches to non-anomalous, associated with
slow-light propagation regime. The observed behavior indicates
the the amplitude of the Stokes field offers an additional
control mechanism of the group index through, for example, the
optical depth of the atomic ensemble. At the same time, the
four-wave mixing process produces higher gain for the probe
field at the output, and thus allows for compensation of
unavoidable optical losses when operating inside a cavity.

\section{Analytical solution}

The results presented above have been obtained by numerical
solution of the propagation equations for all four optical
fields using interaction Hamiltonian, described by
Eq.~(\ref{hamiltonian}) without making any additional
assumptions about the parameters of the system. However, with a
few reasonable approximations, we also can find an analytical
solution for time-dependent weak optical fields $\Omega_2$ and
$\Omega_4$ and strong cw optical fields $\Omega_1$ and
$\Omega_3$. In this case we can assume a linear response of the
atomic medium in response to both weak optical fields. The
strong control fields determine the populations of the atomic
levels and optical polarizations for the $|1\rangle \rightarrow
|3\rangle$ and $|2\rangle \rightarrow |4\rangle$ transitions
that are coupled with these fields. Thus, the corresponding
density matrix elements can be calculated assuming only the
interaction of the two strong fields with the atoms, which in
the interaction scheme under consideration (Fig.~\ref{Nlevels.fig})
reduces to the simple case of two independent two-level
systems, connected only through the decays of the excited
states $|3\rangle$ and $|4\rangle$:
\begin{eqnarray}
\dot{ \rho} _{1,1} &=& \gamma_{31} \rho_{3,3} + \gamma_{41} \rho_{4,4} + \frac{1}{2} i \Omega_1 (\rho _{3,1} - \rho _{1,3}) \label{rho11} \\
\dot{ \rho}_{2,2} &=& \gamma_{32} \rho_{3,3} + \gamma_{42} \rho_{4,4} + \frac{1}{2} i \Omega_2 (\rho _{4,2} - \rho _{2,4}) \\
\dot{ \rho} _{3,3} &=& - \gamma_{3} \rho_{3,3} - \frac{1}{2} i \Omega_1 (\rho _{3,1} - \rho _{1,3}) \\
\dot{ \rho}_{4,4} &=& - \gamma_{4} \rho_{4,4} - \frac{1}{2} i \Omega_2 (\rho _{4,2} - \rho _{2,4}) \\
\dot{ \rho}_{1,3} &=& - (\gamma_{3}/2 + i \delta_1) \rho _{1,3} -\frac{1}{2} i \Omega_1 (\rho _{1,1}-\rho _{3,3}) \\
\dot{ \rho} _{2,4} &=& - (\gamma_{4}/2 + i \delta_3) \rho _{2,4} -\frac{1}{2} i \Omega_3 (\rho _{2,2}-\rho _{4,4}) \label{rho24}
\end{eqnarray}
Here $\gamma_3 = \gamma_{31} + \gamma_{32}$ and $\gamma_4 =
\gamma_{41} + \gamma_{42}$ are the population decay rates of
the excited states. For simplicity, we have neglected the
population decay rates from the two ground states, assuming that
they are significantly smaller than the excited state decays
and the strong optical fields' Rabi frequencies. Comparison with
the exact numerical solutions indicates that this is a good
approximation.

Solving Eqs.~(\ref{rho11}-\ref{rho24}) in the steady state and
assuming equal branching ratios for the excited state decay
channels ($\gamma_{31} = \gamma_{32} = \gamma_3/2$ and
$\gamma_{41} = \gamma_{42} = \gamma_4/2$), we obtain the
following expressions for atomic populations and optical
coherences:
\begin{equation}
\left(
  \begin{array}{c}
    \rho _{1,1}^{(0)} \\
    \rho _{2,2}^{(0)} \\
    \rho _{3,3}^{(0)} \\
    \rho _{4,4}^{(0)} \\
    \rho _{1,3}^{(0)} \\
    \rho _{2,4}^{(0)} \\
  \end{array}
   \right)
  =\frac{1}{T}
  \left(
    \begin{array}{c}
      \Omega_3^2 (4 \delta_1^2 + \Omega_1^2 + \gamma_3^2) \gamma_4\\
      \Omega_1^2 (4 \delta_3^2 + \Omega_3^2 + \gamma_4^2) \gamma_3 \\
      \Omega_1^2 \Omega_3^2 \gamma_4 \\
      \Omega_1^2 \Omega_3^2 \gamma_3  \\
      -\Omega_1 \Omega_3^2 \gamma_4 (2 \delta_1 + i \gamma_3) \\
      -\Omega_3 \Omega_1^2 \gamma_3 (2 \delta_3 + i \gamma_4) \\
    \end{array}
 \right),
\end{equation}
where $T=2 {\Omega_3}^2 \gamma _4 \left(2 {\delta_1}^2+{\Omega_1}^2\right) + \gamma _3 \gamma _4 \left({\Omega_3}^2 \gamma _3+{\Omega_1}^2 \gamma_4\right) +2 {\Omega_1}^2 \gamma _3 \left(2 {\delta_3}^2+ {\Omega_3}^2\right)$ is the common denominator.
We make the additional approximation that the values of these
density matrix elements do not change along the length of the
cell. The validity of this approximation may be questioned,
since, in fact, both strong fields will experience some
absorption. Later, we will demonstrate that in the range of
strong field intensities that produce the desired fast-light
regime this absorption is not significant, and the
non-depletion approximation is reasonable.

We are interested in calculating the propagation of the weak
probe field $\Omega_2$, as well as in the possible generation
of the four-wave mixing field $\Omega_4$ connecting the
$|4\rangle$ and $|1\rangle$ transition, governed by the wave
equation,
\begin{eqnarray} \label{omega24_class}
(-i\omega + c\partial_z)\Omega_2 &=& ig_2N \rho_{3,2}, \label{probe1}\\
(-i\omega + c\partial_z)\Omega_4 &=& ig_4N \rho_{4,1}, \label{stokes1}
\end{eqnarray}
where $g_{2,4}$ are coupling coefficients for the corresponding optical transitions.

The remaining density matrix elements are described by the following equations:
\begin{widetext}
\begin{eqnarray}
\dot{\rho }_{1,2} &=& -\Gamma_{12} \rho _{1,2} + \frac{1}{2} i \Omega_1 \rho_{3,2} -\frac{1}{2} i \Omega_3 \rho_{1,4} - \frac{1}{2} i \Omega_2 \rho _{1,3}^{(0)} + \frac{1}{2} i\Omega _4 \rho _{4,2}^{(0)}; \label{rho12} \\
\dot{\rho }_{1,4} &=& -\Gamma_{14} \rho _{1,4} + \frac{1}{2} i \Omega_1 \rho_{3,4} -\frac{1}{2} i \Omega_3 \rho_{1,2} + \frac{1}{2} i\Omega _4 (\rho _{4,4}^{(0)}-\rho _{1,1}^{(0)}); \label{rho14} \\
\dot{\rho }_{3,2} &=& -\Gamma_{32} \rho _{3,2} + \frac{1}{2} i \Omega_1 \rho_{1,2} -\frac{1}{2} i \Omega_3 \rho_{3,4} - \frac{1}{2} i\Omega _2 (\rho _{3,3}^{(0)}-\rho _{2,2}^{(0)}); \label{rho32} \\
\dot{\rho }_{3,4} &=& -\Gamma_{34} \rho _{3,4} + \frac{1}{2} i \Omega_1 \rho_{1,4} -\frac{1}{2} i \Omega_3 \rho_{3,2} + \frac{1}{2} i \Omega_2 \rho _{2,4}^{(0)} - \frac{1}{2} i\Omega _4 \rho _{3,1}^{(0)}; \label{rho34}
\end{eqnarray}
\end{widetext}
where $\Gamma_{12}=i(\delta_1-\delta_2)$, $\Gamma_{14}=\gamma_4/2 + i(\delta_1-\delta_2+\delta_3)$, $\Gamma_{32}=\gamma_3/2- i\delta_2$, and $\Gamma_{34}=(\gamma_3+\gamma_4)/2 + i(\delta_3-\delta_2)$.

It is important to note that we assume that the detuning of
this generated field is such that it always obeys the
four-photon resonance condition $-\delta_1+\delta_2
-\delta_3+\delta_4=0$. For example, if both strong fields are
tuned to the atomic transition frequencies ($\delta_{1}=\delta_{3}=0$)
and the probe field detuning $\delta_2$ is scanned, the
detuning of the generated Stokes field changes in the opposite
direction $\delta_4=-\delta_2$ to maintain the resonance.

Equations (\ref{rho12}--\ref{rho34}) can be compactly written
as
\begin{equation} \label{densmatr}
\dot{\rho}_\downarrow=M\rho_\downarrow + B,
\end{equation}
where vector $\rho_\downarrow$ consists of the four unknown
density matrix elements
$(\rho_\downarrow)^\mathrm{T}=\{\rho_{1,2}, \rho_{1,4},
\rho_{3,2}, \rho_{3,4}\}$, $M$ is a $4\times4$ matrix:
\begin{equation}
M=\left(
    \begin{array}{cccc}
      i\delta_2 & -i\Omega_3/2 & i\Omega_1/2 & 0 \\
      -i\Omega_3/2 & i\delta_2-\gamma_4/2 & 0 & i\Omega_1/2 \\
      i\Omega_1/2 & 0 & i\delta_2-\gamma_3/2 & -i\Omega_3/2 \\
      0 & i\Omega_1/2 & -i\Omega_3/2 & i\delta_2-\gamma_3/2-\gamma_4/2 \\
    \end{array}
  \right),
\end{equation}
and $B$ is defined as
\begin{equation}
B=\frac{1}{iT}\left(
    \begin{array}{c}
      \Omega_1 \Omega_3 (\Omega_2 \Omega_3+\Omega_1 \Omega_4) \gamma _3 \gamma _4 \\
      i {\Omega_3}^2 \Omega_4 \left({\Omega_1}^2 \gamma _3-{\Omega_1}^2 \gamma _4-\gamma _3^2 \gamma _4\right) \\
      i {\Omega_1}^2 \Omega_2 \left({\Omega_3}^2 \gamma _4-{\Omega_3}^2 \gamma _3-\gamma _3 \gamma _4^2\right) \\
     \Omega_1 \Omega_3 (\Omega_1 \Omega_2+\Omega_3 \Omega_4) \gamma _3 \gamma _4 \\
    \end{array}
  \right).
\end{equation}

In this case the solution of Eq.~(\ref{densmatr}) in the
frequency domain is
 \begin{equation} \label{densmatr}
\rho_\downarrow^{(1)}=-(M+i\omega \mathbf{I})^{-1}B,
\end{equation}
where $\mathbf{I}$ is the identity matrix. Finally, the
calculated expressions for the density matrix elements
$\rho_{3,2}$ and $\rho_{1,4}$ in terms of the optical-field
Rabi frequencies must be substituted into
Eqs.~(\ref{probe1},\ref{stokes1}) to obtain the propagation
equations for the probe and Stokes field in a self-consistent
form:
\begin{equation} \label{omega24}
\partial_z\left(
            \begin{array}{c}
              \Omega_2 \\
              \Omega_4 \\
            \end{array}
          \right) = \frac{iNg}{c}
    M_2 \left(
            \begin{array}{c}
              \Omega_2 \\
              \Omega_4 \\
            \end{array}
          \right),
 \end{equation}
 where the matrix $M_2$ contains the information about atomic response, and we assume equal coupling coefficients $g_2=g_4=g$.
 %
The explicit form of the matrix $M_2$ consists of algebraic
combinations of the Rabi frequencies and detunings of the
strong optical fields and optical transition decay rates, but
is omitted here for brevity.

The important consequence of the non-depletion approximation
for the strong fields is that the right-hand side of
Eq.~(\ref{omega24}) does not depend on position $z$, allowing a
direct solution:

\begin{eqnarray} \label{omega24_2}
\left(
            \begin{array}{c}
              \Omega_2(\omega,z) \\
              \Omega_4(\omega,z) \\
            \end{array}
          \right) &=& e^{\frac{iNg}{c}
            M_2 z} \left(
            \begin{array}{c}
              \Omega_2 (\omega,0) \\
              \Omega_4 (\omega,0)\\
            \end{array}
          \right) \\ \nonumber &\equiv&
          \left(
            \begin{array}{cc}
              A(\omega, z) & B(\omega, z) \\
              C(\omega, z) & D (\omega, z) \\
            \end{array}
          \right)\left(
            \begin{array}{c}
              \Omega_2 (\omega,0) \\
              \Omega_4 (\omega,0)\\
            \end{array}
          \right).
 \end{eqnarray}
 
Here $\Omega_{2,4}(0)$ are the Rabi frequencies corresponding
to the input probe and Stokes fields. It is important to note that expanding the expressions for the coefficients $A$-$D$ forms in
Taylor series up to the $\omega^2$ terms accurately captures the pulse propagation dynamics, but allows significant speed-up in the calculations. The results presented below were obtained in this approximation.

\begin{figure}[htp]
\includegraphics[width=1.0\columnwidth]{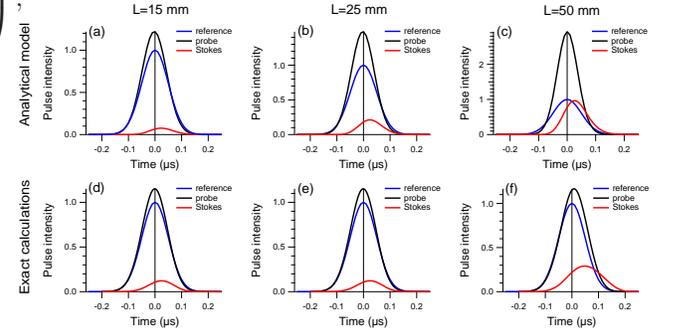}%
\caption{Comparison between exact solution (top) and approximate analytical calculations (bottom) of the signal pulse propagation through the cell of varying length.  }
\label{CalcvsAnalcomp.fig}%
\end{figure}

Fourier transformation of this solution describes the
propagation dynamics of signal/Stokes optical fields.
Fig.~\ref{CalcvsAnalcomp.fig} demonstrates the comparison
between the exact numerical solutions obtained by calculating
all time-dependent density matrix elements and propagation for
all four optical fields, and the prediction of our simplified
analytical theory for propagation of a 100-ns Gaussian probe pulse through an atomic medium with density $10^9~\mathrm{cm}^{-3}$. We observe that, for short lengths of the
atomic medium (15~mm and 25~mm), the two methods provide
similar solutions, predicting small gain and some advance for
the probe optical field, as well as generation of the Stokes
field in a slow-light regime. For the longer cell (50~mm),
however, the analytical model significantly overestimates the
gain in both probe and Stokes fields compared to the exact
numerical solution that takes into the account the attenuation
of both strong control fields associated population
redistribution. Nevertheless, it is interesting to note that
both models predict positive delay for the probe pulse for the
longer cell, with similar delay time.

\begin{figure}[htp]
\includegraphics[width=1.0\columnwidth]{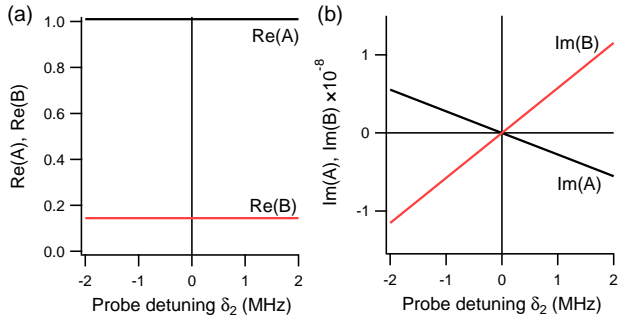}
\caption{Coefficients $A$ and $B$ of the transfer matrix
Eq.~(\ref{omega24_2}) for near-zero probe detuning $\delta_2$.
The calculations are made for conditions identical to those of
Fig.~\ref{CalcvsAnalcomp.fig}(a). }
\label{ABcoef.fig}%
\end{figure}

The analytical solution also provides useful intuition about
the role of the generated Stokes field in the dynamics of the
probe optical field. For example, Fig.~\ref{ABcoef.fig} shows
the real and imaginary parts of the coefficients $A$ and $B$ of
the transfer matrix in Eq.~(\ref{omega24_2}) for a relatively
short atomic medium ($L=1$~cm). The real part of these
coefficients [Fig.~\ref{ABcoef.fig}(a)] illustrates that both
input probe and Stokes fields directly contribute to the
predicted amplification of the probe field after the cell, and
have no spectral dependence near the resonance. The imaginary
parts of the coefficients, shown in Fig.~\ref{ABcoef.fig}(b),
represent the dispersive effect of the atomic medium. They are
both nearly linear functions of frequency, with slopes of
opposite sign. Also, for the chosen detunings,
$\partial\text{Im}(B)/\partial \omega$, representing the Stokes
field contribution to the dispersion, is approximately twice as
steep as $\partial\text{Im}(A)/\partial \omega$. Thus, it is
not surprising that for very weak Stokes fields (corresponding
to low optical depth values) the dispersion is predominantly
determined by the probe field propagation, and displays ``fast
light'' regime. As the amplitude of the Stokes field increases,
it adds up with the opposite phase to the output field, and,
eventually, changes the sign of the dispersion. Under these
conditions, the output probe field is delayed, as in the ``slow
light'' regime.

\section{Conclusions}

In conclusion, we have analyzed the propagation of a weak resonant probe
through a medium of four-level atoms in an \emph{N}-scheme with allowed
four-wave mixing generation, and found it to be a promising
candidate for the realization of tunable ``slow-to-fast'' light
with no absorption. This is particularly interesting
for the experimental investigation of potential techniques for
the enhancement of optical-gyroscope performance.

\section{Acknowledgments}

The authors thank Frank Narducci and John Davis for useful discussions. This
research was supported by Naval Air Warfare Center STTR program
N68335-11-C-0428.



\end{document}